\documentclass[conference]{IEEEtran}
\IEEEoverridecommandlockouts

\usepackage{cite}
\usepackage{amsmath,amssymb,amsfonts}
\usepackage{algorithmic}
\usepackage{graphicx}
\usepackage{textcomp}
\usepackage{xcolor}

\usepackage{booktabs} 
\usepackage{url}
\usepackage{listings}
\usepackage[frozencache=true]{minted}
\usepackage{threeparttable}
\usepackage{subcaption}
\usepackage{enumitem}
\usepackage[binary-units=true]{siunitx}
\usepackage{ragged2e}
\usepackage[altpo]{backnaur}
\usepackage{nccmath}
\usepackage{bigstrut}
\usepackage{sepnum}
\usepackage{mathpartir}
\usepackage[linesnumbered,ruled]{algorithm2e}
\def\BibTeX{{\rm B\kern-.05em{\sc i\kern-.025em b}\kern-.08em
    T\kern-.1667em\lower.7ex\hbox{E}\kern-.125emX}}

\newcommand{\nnum}[1]{\sepnum{.}{,}{}{#1}}
\PassOptionsToPackage{hyphens}{url}\usepackage{hyperref}

\setlist[itemize]{noitemsep, leftmargin=1em}
\setlist[enumerate]{noitemsep, leftmargin=1em}

\SetCommentSty{mycommfont}

\newcommand{\prim}[1]{\text{{\sffamily{\small {\bf #1}}}}}
\newcommand{\primsem}[1]{\text{{\sffamily{\footnotesize {\bf #1}}}}}

\newcommand{\pycg}{{\sc p}y{\sc cg}}

\begin{document}
\title{PyCG: Practical Call Graph Generation in Python}

\author{\IEEEauthorblockN{Vitalis Salis,\textsuperscript{$\mathsection$}\textsuperscript{$\dagger$} Thodoris Sotiropoulos,\textsuperscript{$\mathsection$}
Panos Louridas,\textsuperscript{$\mathsection$} Diomidis Spinellis\textsuperscript{$\mathsection$}
and Dimitris Mitropoulos\textsuperscript{$\mathsection$}\textsuperscript{$\ddagger$}}
\IEEEauthorblockA{\textsuperscript{$\mathsection$}\textit{Athens University of Economics and Business}\\
\textsuperscript{$\dagger$}\textit{National Technical University of Athens}\\
\textsuperscript{$\ddagger$}\textit{National Infrastructures for Research and Technology - GRNET}\\
vitsalis@gmail.com, \{theosotr, louridas, dds, dimitro\}@aueb.gr}
}

\maketitle

\thispagestyle{plain}
\pagestyle{plain}

\begin{abstract}
Call graphs play an important role
in different contexts, such as
profiling and
vulnerability propagation analysis.
Generating call graphs in an efficient manner
can be a challenging task
when it comes to high-level
languages that are modular
and incorporate
dynamic features and higher-order functions.

Despite the language's popularity,
there have been very few tools
aiming to generate call graphs for Python programs.
Worse,
these tools suffer from several effectiveness issues
that limit their practicality in
realistic programs.
We propose a pragmatic,
static approach for call graph
generation in Python.
We compute all assignment relations
between program identifiers of
functions, variables, classes, and modules
through an inter-procedural analysis.
Based on these assignment relations,
we produce the resulting call graph
by resolving all calls to
potentially invoked functions.
Notably,
the underlying analysis is designed to be efficient
and scalable,
handling several Python features,
such as modules,
generators,
function closures,
and multiple inheritance.

We have evaluated our prototype implementation,
which we call PyCG,
using two benchmarks:
a micro-benchmark suite containing
small Python programs
and a set of macro-benchmarks with several
popular real-world Python packages.
Our results indicate that PyCG can
efficiently handle thousands of lines of
code in less than a second
(0.38 seconds for 1k LoC on average).
Further,
it outperforms the state-of-the-art for Python
in both precision and recall:
PyCG achieves high rates of precision $\sim$99.2\%,
and adequate recall $\sim$69.9\%.
Finally,
we demonstrate how PyCG can aid
dependency impact analysis by showcasing
a potential enhancement to
GitHub's ``security advisory''
notification service using a real-world example.
\end{abstract}


\section{Introduction}
\label{sec:intro}

A call graph depicts calling
relationships between subroutines in a computer program.
Call graphs can be employed
to perform a variety of tasks, such as
profiling~\cite{callgrind},
vulnerability propagation~\cite{SZ12},
and tool-supported refactoring~\cite{FMMS11}.

Generating call graphs in an efficient way
can be a complex endeavor especially
when it comes to high-level,
dynamic programming languages.
Indeed,
to create precise call graphs for programs
written in languages such as Python and JavaScript,
one must deal with several challenges
including higher-order functions,
dynamic and metaprogramming features (e.g., {\tt eval}),
and modules.
Addressing such challenges can play a
significant role in the improvement of
dependency impact analysis~\cite{HDG18, KGDP17, npmblog},
especially in the context of package managers
such as {\it npm}~\cite{npm}
and {\it pip}~\cite{pip}.

To support call graph generation
in dynamic languages,
researchers have proposed different
methods relying on static analysis.
The primary aim for many
implementations is completeness,
i.e., facts deduced by the system are indeed
true~\cite{JMT09,LWJC12,KDKW14}.
However,
for dynamic languages,
completeness comes with a performance cost.
Hence,
such approaches
are rarely employed in practice
due to scalability issues~\cite{KLDR15}.
This has led to
the emergence of {\it practical} approaches
focusing on incomplete static analysis
for achieving better performance~\cite{MLF13,FSSD13}.
Sacrificing completeness is the key enabler
for adopting these approaches
in applications
that interact with complex libraries~\cite{MLF13},
or Integrated Development Environments ({\sc ide}s)~\cite{FSSD13}.
Prior work
primarily targets JavaScript programs and---%
among other things---attempts to address challenges
related to events and the language's
asynchronous nature~\cite{async,event}.

Despite Python's popularity~\cite{python-github}, there have been
surprisingly few tools aiming to generate call graphs for programs
written in the language. {\it Pyan}~\cite{FHJM18} parses the program's
Abstract Syntax Tree ({\sc ast}) to extract its call graph.
Nevertheless, it has drawbacks in the way it handles the
inter-procedural flow of values and module imports.
{\it code2graph}~\cite{GTL18, GAL18} visualizes {\it Pyan}-constructed call
graphs, so it has the same limitations. {\it Depends}~\cite{DEP18}
infers syntactical relations among source code entities to generate
call graphs. However, functions assigned to variables or passed to
other functions are not handled by {\it Depends}, thus it does not
perform well in the context of a language supporting higher-order
programming. We will expand on the shortcomings of the existing tools
in the remainder of this work. That said, developing an effective and
efficient call graph generator for a dynamically typed language like
Python is no minor task.

We introduce a practical approach
for generating call graphs for 
Python programs and implement a
corresponding prototype that
we call \pycg.
Our approach works in two steps.
In the first step
we compute the~\emph{assignment graph},
a structure that shows
the assignment relations
among program identifiers.
To do so,
we design a context-insensitive
inter-procedural analysis
operating on a simple intermediate representation
targeted for Python.
Contrary to the existing static analyzers,
our analysis is capable of handling
intricate Python features,
such as higher-order functions,
modules,
function closures,
and multiple inheritance.
In the next step,
we build the call graph
of the original program
using the assignment graph.
Specifically,
we utilize the graph
to resolve all functions
that can be potentially pointed to by
callee variables.
Such a programming pattern
is particularly common
in higher-order programming.

Similar to previous work~\cite{FSSD13},
our analysis follows a conservative approach,
meaning that the analysis does not reason about
loops and conditionals.
To make our analysis more precise,
especially when dealing with features like
inheritance, modules
or programming patterns such as duck typing~\cite{duck},
we distinguish attribute accesses (i.e, $e.x$)
based on the namespace 
where the attribute ($x$) is defined.
Prior work uses a~\emph{field-based} approach
that correlates attributes of the same name
with a single global location
without taking into account
their namespace~\cite{FSSD13}.
This leads to false positives.
Our design choices make our approach achieve
high rates of precision,
while remaining efficient
and applicable to large-scale Python programs.

We evaluate the effectiveness of our method
through a micro- and a macro-benchmarking suite.
Also,
we compare it against
{\it Pyan} and {\it Depends}.
Our results indicate that our method
achieves high levels of precision ($\sim$99.2\%)
and adequate recall ($\sim$69.9\%)
on average,
while the other analyzers
demonstrate lower rates in both measures.
Our method is able to handle
medium-sized projects in less than one second
(0.38 seconds for 1k LoC on average).
Finally,
we show how our method can accommodate
the fine-grained tracking of vulnerable dependencies
through a real-world case study.

\noindent
{\bf Contributions.}
Our work makes the following contributions.
\begin{itemize}
\item We propose a static approach for
pragmatic call graph generation in Python.
Our method performs inter-procedural analysis
on an intermediate language that
records the assignment relations between
program identifiers, i.e., functions,
variables,
classes and modules.
Then it examines the documented associations
to extract the call graph
(Section~\ref{sec:approach}).
\item We develop a micro-benchmark suite
that can be used as a standard to evaluate
call graph generation methods in Python.
Our suite is modular,
easily extendable, and covers a large fraction
of Python's functionality related to classes,
generators,
dictionaries, and more
(Section~\ref{sec:evaluation:micro-benchmark}).
\item We evaluate the effectiveness of our approach
through our micro-benchmark and
a set of macro-benchmarks including
several medium-sized Python projects.
In all cases our method achieves high rates
of precision and recall,
outperforming the other available analyzers
(Sections~\ref{sec:evaluation:micro}, \ref{sec:evaluation:macro}).
\item We demonstrate how our approach can
aid dependency impact analysis
through a potential enhancement
of GitHub's ``security advisory''
notification service
(Section~\ref{sec:github}).
\end{itemize}

\noindent
\\{\bf Availability.}
PyCG is available as open-source software under the Apache 2.0 Licence at
\url{https://github.com/vitsalis/pycg}. The research artifact is available at
\url{https://doi.org/10.5281/zenodo.4456583}.

\section{Background}
\label{sec:background}
Generating precise call graphs for Python programs
involves several challenges.
Existing static approaches fail
to address these challenges
leaving opportunities for improvement.

\subsection{Challenges}
\label{sec:background:challenges}

\begin{itemize}
\item {\it Higher-order Functions}:
In a high-level language such as Python,
functions can be assigned to variables,
passed as parameters to other functions,
or serve as return values.
\item {\it Nested Definitions}:
Function definitions can be nested,
meaning that a function
can be defined and invoked within the context
of another function.
\item {\it Classes}:
As an object-oriented language,
Python allows for the creation of classes
that inherit attributes and methods from other classes.
The resolution of inherited methods from
parent classes requires the computation
of the Method Resolution Order ({\sc mro})
of each class.
\item {\it Modules}:
Python is highly extensible,
allowing applications to import different modules.
Keeping track of the different
modules that are imported in an application,
as well as the resolution order of those imports,
can be a challenging task.
\item {\it Dynamic Features}:
Python is dynamically typed,
allowing variables to take values
of different types during execution.
Also,
it allows for classes to be
dynamically modified during runtime.
Furthermore,
the {\tt eval} function
allows for a dynamically constructed string
to be executed as code.
\item {\it Duck Typing}:
Duck typing is a programming pattern
that is particularly common
in dynamic languages such as Python~\cite{duck}.
Through duck typing,
the suitability of an object
is determined by the presence of specific
methods and properties,
rather than the type of the object itself.
In this context,
given a method defined by two
(or more) classes,
it is not trivial to identify
its origins when it is invoked.
\end{itemize}

\subsection{Limitations of Existing Static Approaches}
\label{sec:background:limitations}

\begin{figure}[!t]
\begin{minted}[fontsize=\footnotesize,linenos,xleftmargin=20pt,breaklines]{python}
import cryptops

class Crypto:
    def __init__(self, key):
        self.key = key

    def apply(self, msg, func):
        return func(self.key, msg)

crp = Crypto("secretkey")
encrypted = crp.apply("hello world", cryptops.encrypt)
decrypted = crp.apply(encrypted, cryptops.decrypt)
\end{minted}
\vspace{-1mm}
\caption{The {\tt crypto} module.
Existing tools fail to generate
a corresponding call graph effectively.}
\label{fig:crypto-mod}
\vspace{-5mm}
\end{figure}

\begin{figure*}[!t]
\centering
\begin{subfigure}[t]{0.4\linewidth}
    \centering
    \includegraphics[scale=0.3]{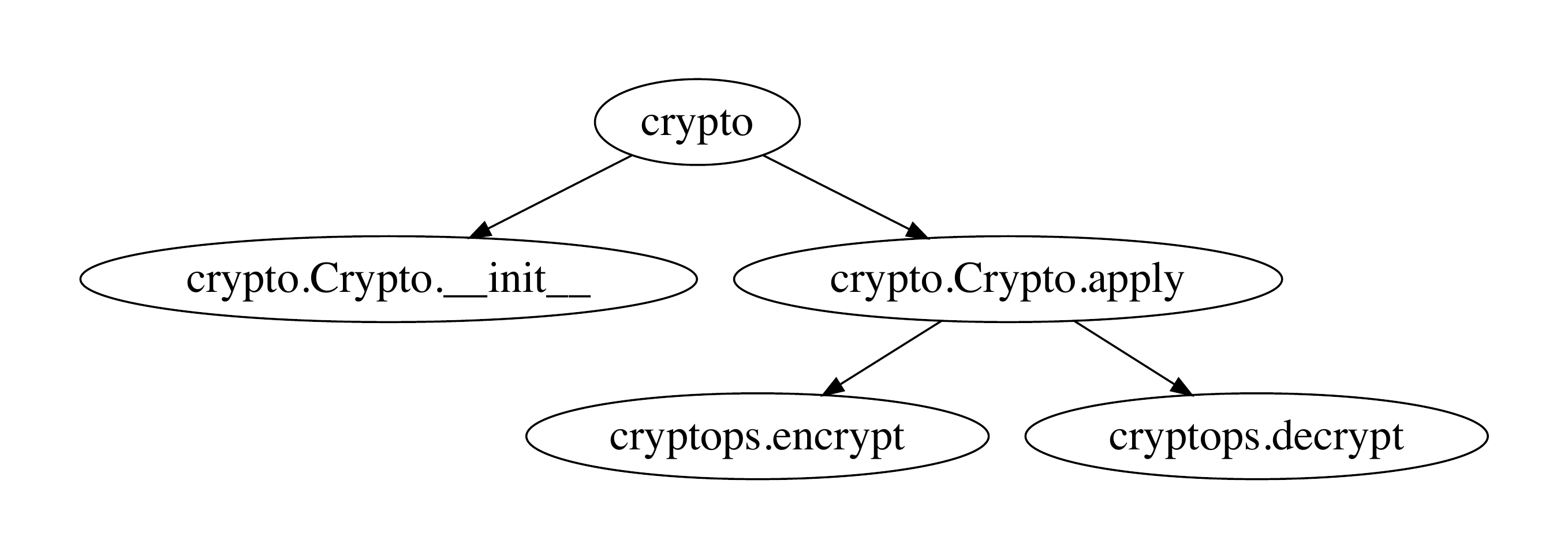}
	\vspace{-1mm}
    \caption{Precise call graph.}
    \label{fig:crypto-cg-correct}
\end{subfigure}
\begin{subfigure}[t]{0.3\linewidth}
    \centering
    \includegraphics[scale=0.3]{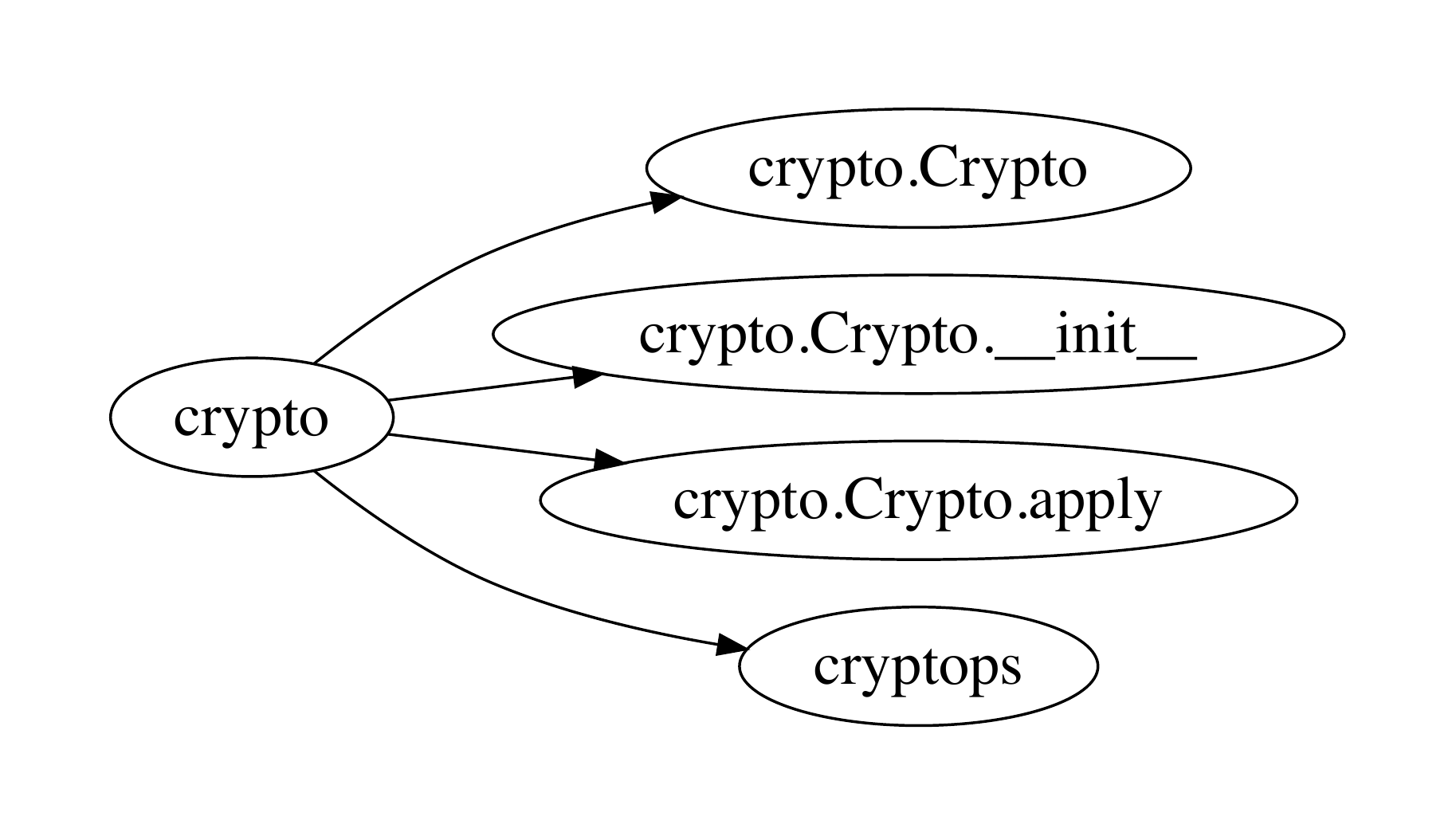}
	\vspace{-1mm}
    \caption{{\it Pyan}-generated call graph.}
    \label{fig:crypto-pyan-cg}
\end{subfigure}
\begin{subfigure}[t]{0.25\linewidth}
    \centering
    \includegraphics[scale=0.3]{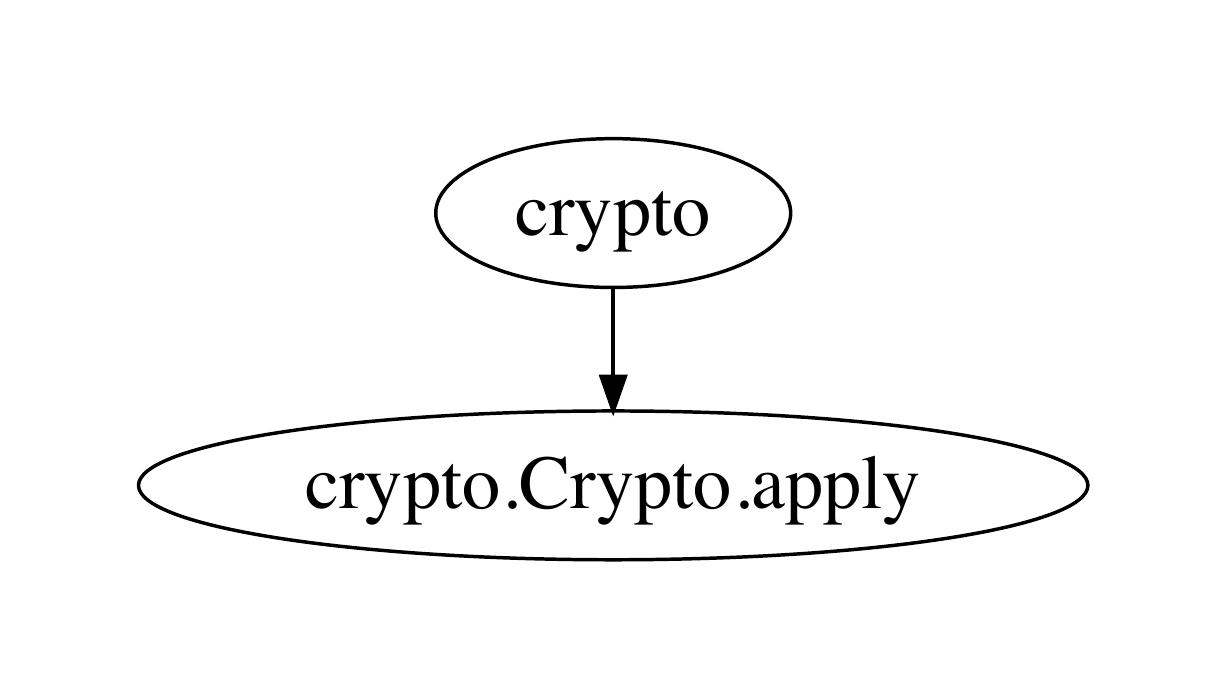}
	\vspace{3mm}
    \caption{{\it Depends}-generated call graph.}
    \label{fig:crypto-depends-cg}
\end{subfigure}
\caption{\label{fig:crypto-cg} Call graphs for the {\tt crypto} module.}
\vspace{-3mm}
\end{figure*}

We focus on two {\it open-source}
static analyzers:
{\it Pyan}~\cite{FHJM18} and
{\it Depends}~\cite{DEP18}.
We do not examine code2graph~\cite{GTL18, GAL18} separately,
as it is based on {\it Pyan} to generate call graphs.
We discuss the limitations of the two existing analyzers
in terms of efficiency and practicality.
To do so,
we introduce a small Python module
named {\tt crypto}
(see~Figure~\ref{fig:crypto-mod}),
which is used
to encrypt and decrypt a
``hello world'' message.
First,
it imports an external Python module named
{\tt cryptops},
which defines two functions,
namely:
{\tt encrypt(key, msg)} and
{\tt decrypt(key, msg)}.
Then, the
{\tt Crypto} class
is defined.
To use it, we instantiate it with an encryption key
and we can encrypt or decrypt messages
by calling {\tt apply(self, msg, func)},
where {\tt func} is one of
{\tt encrypt(key, msg)} and
{\tt decrypt(key, msg)}.
Figure~\ref{fig:crypto-cg-correct} shows
the call graph of the module.

{\it Pyan}~\cite{FHJM18} produces
the imprecise call graph shown in Figure~\ref{fig:crypto-pyan-cg}.
This graph does not
contain all function calls,
because the tool does not track the
inter-procedural flow of values.
Therefore,
it is unable to infer
which functions are passed
as arguments to
{\tt apply(self, msg, func)}.
In addition,
there are several features that
lead to the addition of
unrealized call edges.
Specifically,
when {\it Pyan} detects
object initialization,
it creates call edges to
both the class name and
the {\tt \_\_init\_\_()}
method of the class.\footnote{In
Python,
{\tt \_\_init\_\_()} is the name of a special function
called during object construction.}
Beyond that,
in the case of a module import,
{\it Pyan} generates a call edge
from the importing namespace to
the module name.

{\it Depends}
produces the call graph
presented in Figure~\ref{fig:crypto-depends-cg}.
{\it Depends}
does not track function calls originating from
the module's namespace
(e.g., {\tt crp.apply()}).
This in turn,
led to an empty call graph.
Therefore,
to get a result,
we wrapped those function calls
within a new function.
The resulting graph
does not contain most of the calls
included in the source program.
This is because
{\it Depends} does not capture the call to
the {\tt \_\_init\_\_()} function
of the {\tt Crypto} class.
Furthermore,
(like {\it Pyan})
{\it Depends} does not track the inter-procedural flow
of functions leading to missing edges to
the parameter functions.
Compared to {\it Pyan},
{\it Depends} follows a more conservative approach.
That is,
it only includes a call edge when it has
all the necessary information it needs to
anticipate that the call will be realized.
Contrary to {\it Pyan},
this can lead to a call graph without
false positives.

\vspace{-0.4mm}
\section{Practical Call Graph Generation}
\label{sec:approach}

Our approach for generating call graphs
employs
a context-insensitive inter-procedural analysis
operating on an intermediate representation
of the input Python program.
The analysis uses a fixed-point iteration algorithm,
and gradually builds the~\emph{assignment graph},
which is a structure that shows
the assignment relations between program identifiers
(Section~\ref{sec:analysis}).
In a language supporting higher-order programming,
the assignment graph is an essential component
that we use for resolving functions
pointed to by variables.
Function resolution takes place
at the final step
where we build the call graph for
the given program by
exploiting the assignment graph
stemming from the analysis step
(Section~\ref{sec:cg-construct}).

\begin{figure}[t]
\small
\begin{bnf*}
    \bnfprod*{$e \in {\it Expr}$}
    {\bnftd{o} \bnfor
    \bnftd{x} \bnfor
    \bnftd{x := e} \bnfor
    \bnftd{\prim{function} x (y\dots) e} \bnfor
    \bnftd{\prim{return} e} \bnfor}\\[-0.5mm]
    \bnfmore*{
    \bnftd{e(x=e\dots)} \bnfor
    \bnftd{\prim{class} x (y\dots) e} \bnfor
    \bnftd{e.x} \bnfor
    \bnftd{e.x := e} \bnfor}\\[-0.5mm]
    \bnfmore*{
    \bnftd{\prim{new} x $(y=e\dots)$} \bnfor
    \bnftd{\prim{import} x \prim{from} m \prim{as} y} \bnfor}\\[-0.5mm]
    \bnfmore*{
    \bnftd{\prim{iter} $x$} \bnfor
    \bnftd{e;e}}\\[-0.5mm]
    \bnfprod*{$o \in {\it Obj}$}
    {\bnftd{$n, v$}}\\[-0.5mm]
    \bnfprod*{$v \in {\it Definition}$}
    {\bnftd{$x, \tau$}}\\[-0.5mm]
    \bnfprod*{$\tau \in {\it IdentType}$}
    {\bnftd{\prim{func}} \bnfor
     \bnftd{\prim{var}} \bnfor
     \bnftd{\prim{cls}} \bnfor
     \bnftd{\prim{mod}}}\\[-0.5mm]
    \bnfprod*{$n \in {\it Namespace}$}
    {\bnftd{$(v)^{*}$}}\\[-0.5mm]
    \bnfprod*{$x,y \in {\it Identifier}$}
    {\bnftd{\text{is the set of program identifiers}}}\\[-0.5mm]
    \bnfprod*{$m \in {\it Modules}$}
    {\bnftd{\text{is the set of modules}}}\\[-0.5mm]
    \bnfprod*{$E$}
    {\bnftd{$[]$} \bnfor
     \bnftd{x := E} \bnfor
     \bnftd{\prim{return} E} \bnfor
     \bnftd{$E(x=e\dots)$} \bnfor}\\[-0.5mm]
     \bnfmore*{
     \bnftd{$o(x=E\dots)$} \bnfor
     \bnftd{\prim{new} x(y=E)} \bnfor
     \bnftd{E.x} \bnfor
     \bnftd{E.x := e} \bnfor}\\[-0.5mm]
     \bnfmore*{
     \bnftd{o.x := E} \bnfor
     \bnftd{\prim{iter} $o$} \bnfor
     \bnftd{E;e} \bnfor
     \bnftd{o;E}}
\end{bnf*}
\vspace{-3mm}
\caption{The syntax for representing the input Python programs
along with the evaluation contexts.}\label{fig:syntax}
\vspace{-6.5mm}
\end{figure}

\subsection{The Core Analysis}
\label{sec:analysis}

The starting point of our approach
is to compute the assignment graph
using an inter-procedural analysis
working on an intermediate representation
targeted for Python programs.

One of the key elements of our analysis
is that it examines attribute
accesses based on the namespace
where each attribute is defined.
For example,
consider the following code snippet:
\begin{minted}[fontsize=\footnotesize,linenos,xleftmargin=20pt]{python}
class A:
  def func():
    pass

class B:
  def func():
    pass

a = A()
b = B()
a.func()
b.func()
\end{minted}
Our analysis is able to distinguish
the two functions defined at lines~2 and~6,
because they are members of two different classes,
i.e., class {\tt A} and {\tt B} respectively.
Note that field-based approaches
focused on JavaScript~\cite{FSSD13}
will fail to treat the two invocations as different,
causing imprecision.
That is because a field-based approach
will match all accesses of 
identical attribute names (e.g., {\tt func()}) with
a single object.

\subsubsection{Syntax}
\label{sec:syntax}

The intermediate representation,
where our analysis works on,
follows the syntax 
of a simple imperative
and object-oriented language,
which is shown in Figure~\ref{fig:syntax}.
The last rule in this figure also
shows the evaluation contexts~\cite{semantics}
for this language,
which we will explain shortly.

An important element of this model language
is identifiers.
Every identifier can be one of
the following four types:
(1) \prim{func} corresponding to the name of a function
(2) \prim{var} indicating the name of a variable,
(3) \prim{cls} for class names,
and (4) \prim{mod}
when the identifier is a module name.
Every pair $(x,\tau) \in {\it Identifier} \times {\it IdentType}$
forms a definition.
We represent every definition
and its namespace as an object
(see the ${\it Obj}$ rule).
A namespace is a sequence of definitions,
and it is essential for distinguishing
objects sharing
the same identifier from each other.
For example,
consider the following Python code fragment
located in a module named {\tt main}.
\begin{minted}[fontsize=\footnotesize,linenos,xleftmargin=20pt]{python}
var = 10
class A:
  var = 10
\end{minted}
The analysis distinguishes
the objects created at lines 1 and 3,
as the first one resides
in the namespace $[({\tt main}, \prim{mod})]$,
while the second one lives
in the namespace $[({\tt main}, \prim{mod}), ({\tt A}, \prim{cls})]$.

Our approach treats
every object as the~\emph{value}
given from the evaluation of the expressions
supported by the language.
In particular,
our representation contains expressions
that capture the inter-procedural flow,
assignment statements,
class and function definitions,
module imports,
and iterators / generators
(see the ${\it Expr}$ rule).
Note that the language
is able to abstract
different features,
including lambda expressions,
keyword arguments,
constructors,
multiple inheritance,
and more.

As with prior work
focusing on JavaScript~\cite{event,promises,async},
we use evaluation contexts~\cite{semantics}
that describe the order in which
sub-expressions are evaluated.
For example,
in an attribute assignment $E.x := e$,
the $E$ symbol denotes
that we are currently evaluating the receiver
of the attribute $x$,
while $o.x := E$ indicates
that the receiver has been already evaluated
to an object $o \in {\it Obj}$
(recall that evaluating expressions results
in objects),
and the evaluation now proceeds to the right-hand side
of the assignment.

{\bf Remarks.}
When calling Python functions that produce a generator
(i.e., they contain a {\tt yield} statement
instead of {\tt return}),
these calls take place only when
the generator is actually used.
To model this effect,
when encountering such lazy calls
(e.g., {\tt gen = lazy\_call(x)}),
we create a thunk
(e.g., {\tt gen = lambda: lazy\_call(x)})
that is evaluated only
when we iterate the generator
(through the~\prim{iter} construct).
Furthermore,
dictionaries and lists are treated as regular objects.
For example,
we model a dictionary lookup {\tt x["key"]},
as an attribute access $x.key$.

\subsubsection{State}

\begin{figure}[t]
\centering
\small
\begin{align*}
    &\pi \in {\it AssignG}    = {\it Obj} \hookrightarrow \mathcal{P}({\it Obj}) \\
    &s \in {\it Scope}        = {\it Definition} \hookrightarrow \mathcal{P}({\it Definition}) \\
    &h \in {\it ClassHier}    = {\it Obj} \hookrightarrow {\it Obj}^{*} \\
    &\sigma \in {\it State}   = {\it AssignG} \times {\it Scope} \times {\it Namespace} \times {\it ClassHier}
\end{align*}
\vspace{-6mm}
\caption{Domains of the analysis.}\label{fig:domains}
\vspace{-6mm}
\end{figure}

After converting the original Python program
to our intermediate representation,
our analysis starts evaluating each expression,
and gradually constructs the assignment graph.
To do so,
the analysis maintains a state
consisting of four domains
as shown in Figure~\ref{fig:domains},
namely,
\emph{scope},
\emph{class hierarchy},
\emph{assignment graph},
and~\emph{current namespace}.

\begin{figure*}[th]
\centering
\begin{subfigure}[t]{0.45\linewidth}
\centering
\includegraphics[scale=0.28]{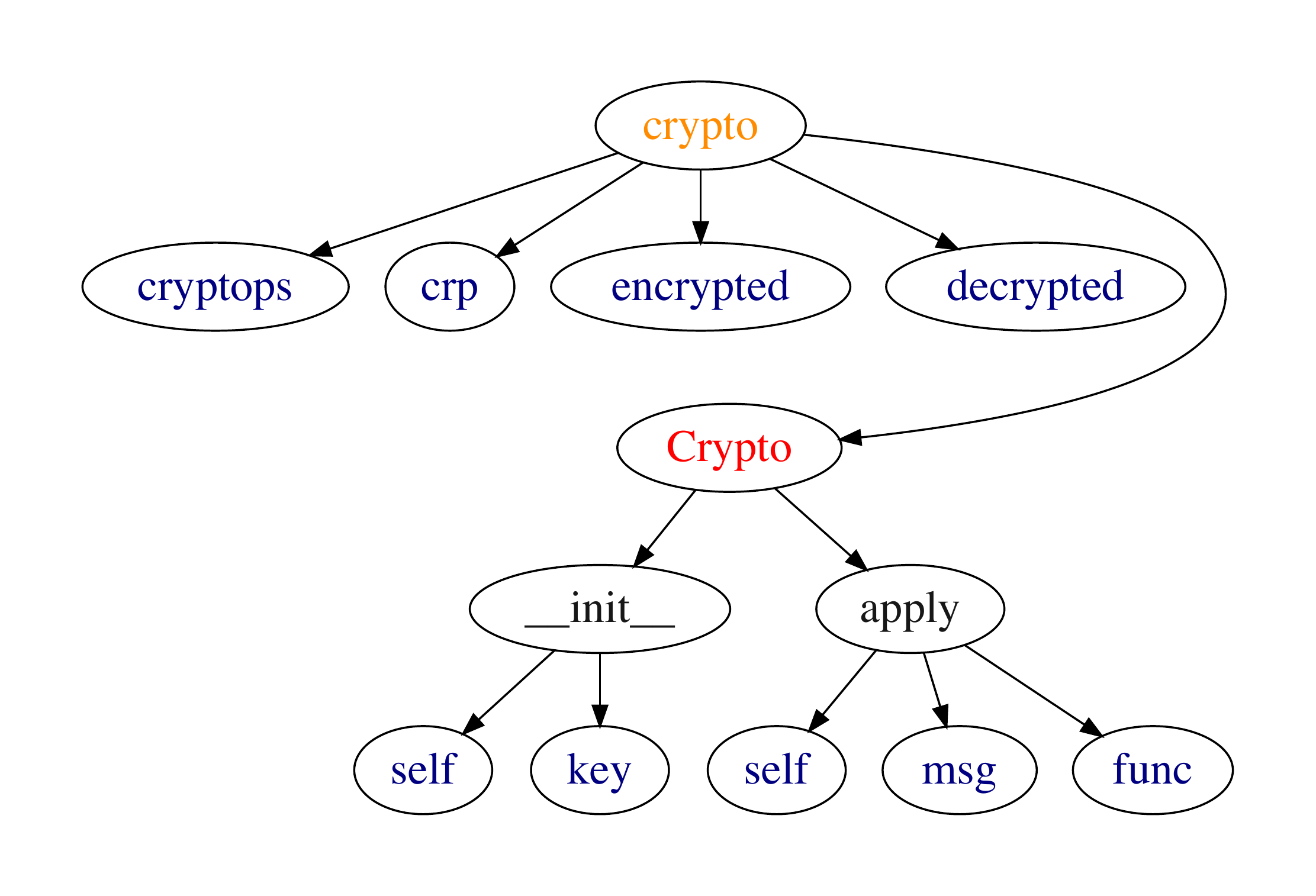}
\vspace{-3.4mm}
\caption{The scope tree of the {\tt crypto} module.}
\label{fig:scope_tree}
\end{subfigure}
\begin{subfigure}[t]{0.45\linewidth}
    \centering
    \includegraphics[scale=0.28]{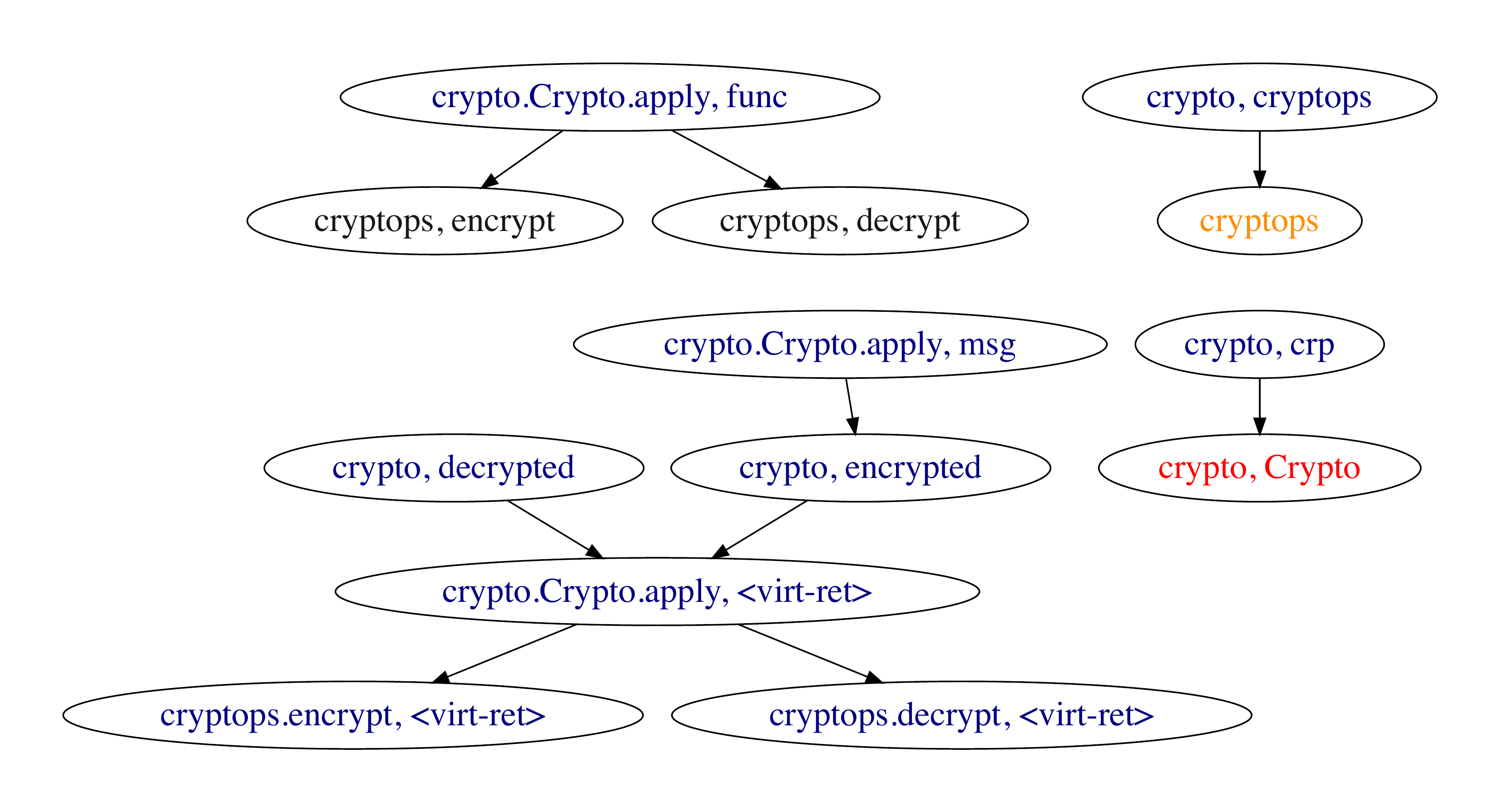}
	\vspace{-4.4mm}
    \caption{The assignment graph of the {\tt crypto} module.}
    \label{fig:ass_graph}
\end{subfigure}
\vspace{-1mm}
\caption{\label{fig:scope-ass} Analyzing the {\tt crypto} module.}
\vspace{-5mm}
\end{figure*}

A scope is a map of definitions
to a set of definitions.
Conceptually,
a scope is a tree
where each node corresponds to a definition
(e.g., a function),
and each edge shows the parent/child relations
between definitions,
i.e., the target node is defined
inside the definition of the source node.
The domain of scopes is useful
for correctly resolving the definitions
that are visible inside
a specific namespace.
Figure~\ref{fig:scope_tree} illustrates
the scope tree of the program depicted
in Figure~\ref{fig:crypto-mod},
and shows all program definitions
and their inter-relations.
Orange nodes correspond to module definitions,
red nodes are class definitions,
black nodes indicate functions,
while blue nodes denote variables.
Based on this scope tree,
we infer that the function {\tt apply} is defined
inside the class {\tt Crypto},
which is in turn defined
inside the module {\tt crypto},
i.e.,
notice the path
{\tt crypto} $\rightarrow$ {\tt Crypto} $\rightarrow$ {\tt apply}.
This domain enables us
to properly deal with Python features
such as function closures
and nested definitions.

A class hierarchy is a tree
representing the inheritance relations
among classes.
An edge from node $u$ to node $v$
indicates that the class $u$ is a child
of the class $v$.
The analysis uses 
this domain for
resolving class attributes
(either methods or fields)
defined in the base classes
of the receiver object.
Through this domain 
we are able to handle
the object-oriented nature of Python,
addressing features such as
multiple inheritance,
and the method resolution order.

The assignment graph is defined as a map
of objects to an element of
the power set of objects $\mathcal{P}({\it Obj})$.
This graph holds
the assignment relations between objects,
capturing the assignments
and the inter-procedural flow of the program.
Figure~\ref{fig:ass_graph} illustrates
the assignment graph corresponding to
the program of Figure~\ref{fig:crypto-mod}.
Each node in the graph
(e.g., {\tt \{crypto.Crypto.apply, func\}})
represents an object.
The first component of the node label
(e.g., {\tt crypt.Crypto.apply})
indicates the namespace where
each identifier (e.g., {\tt func}) is defined.
Colors reveal the type of the identifier
as explained in a previous paragraph
(e.g., the blue color implies variable definitions).
An edge shows the possible values
that a variable may hold.
For example,
the variable {\tt func} defined
in the {\tt crypto.Crypto.apply} namespace
may point to
the functions {\tt decrypt}
and {\tt encrypt},
both defined in the {\tt cryptops} namespace.
As another example,
notice the edge
originating from the node {\tt \{crypto.Crypto.apply, msg\}}
and leading to {\tt \{crypto, encrypted\}}.
This edge shows that the parameter {\tt msg}
of the function {\tt crypto.Crypto.apply}
points to the variable {\tt encrypted}
when the function is invoked on line 12.
The assignment graph domain
enables us to address the challenge regarding
higher-order programming in Python.

Finally,
we use the current namespace
to track the location
where new variables,
classes, modules, and functions are defined.
This domain is important
for establishing a more precise analysis
than field-based analysis employed by prior work.
Through namespaces,
objects and attribute accesses
are distinguished based on their namespace,
addressing challenges such as duck typing.

\subsubsection{Analysis Rules}

\begin{figure}
\centering
\footnotesize
\resizebox{0.9\linewidth}{!}{%
\begin{mathpar}
\inferrule[e-ctx]{
    \langle\pi, s, n, h, e\rangle \hookrightarrow \langle\pi', s', n', h', e'\rangle
}
{\langle\pi, s, n, h, E[e]\rangle \rightarrow \langle\pi', s', n', h', E[e']\rangle}
\hva \and
\inferrule[compound]{
}
{\langle\pi, s, n, h, E[o_1;o_2]\rangle \rightarrow \langle\pi, s, n, h, E[o_2]\rangle}
\hva \and
\inferrule[ident]{
  o = {\tt getObject}(s, n, x)
}
{\langle\pi, s, n, h, E[x]\rangle \rightarrow \langle\pi, s, n, h, E[o]\rangle}
\hva \and
\inferrule[assign]{
    s' = {\tt addScope}(s, n, x, \primsem{var}) \\
    o' = (n, (x, \primsem{var})) \\
  \pi' = \pi[o' \rightarrow \pi(o') \cup \{o\}] \\
}
{\langle\pi, s, n, h, E[x := o]\rangle \rightarrow \langle\pi', s', n, h, E[o']\rangle}
\hva \and
\inferrule[func]{
    s' = {\tt addScope}(s, n, x, \primsem{func}) \\
    n' = n \cdot (x, \primsem{func}) \\
    s''= {\tt addScope}(s', n', {\tt ret}, \primsem{var}) \\
    s^{(3)} = {\tt addScope}(s'', n', y, \primsem{var}) \\
}
{\langle\pi, s, n, h, E[\text{\primsem{function} $x$ $(y\dots)$ $e$}]\rangle \rightarrow \langle\pi, s^{(3)}, n', h, E[e]\rangle}
\hva \and
\inferrule[return]{
    o' = (n \cdot x, ({\tt ret}, \primsem{var})) \\
  \pi' = \pi[o' \rightarrow \pi(o') \cup \{o\}] \\
}
{\langle\pi, s, n \cdot x, h, E[\text{\primsem{return} $o$}]\rangle \rightarrow \langle\pi', s, n, h, E[o']\rangle}
\hva \and
\inferrule[call]{
    o_1 = (n', (f, \primsem{func})) \\
    o_2' = (n' \cdot f, (y, \primsem{var})) \\
   \pi' = \pi[o_2' \rightarrow \pi(o_2') \cup \{o_2\}]
}
{\langle\pi, s, n, h, E[o_1(y=o_2\dots)]\rangle \rightarrow \langle\pi', s, n, h, (n' \cdot f, ({\tt ret}, \primsem{var}))\rangle}
\hva \and
\inferrule[class]{
    s' = {\tt addScope}(s, n, x, \primsem{cls}) \\
    t = \langle{\tt getObject(s, n, b)} \mid b \in (y\dots)\rangle \\
  h' = h[(n, (x, \primsem{cls})) \rightarrow t] \\\
  n' = n \cdot (x, \primsem{cls})
}
{\langle\pi, s, n, h, E[\text{\primsem{class} $x$ ($y\dots$) $e$}]\rangle \rightarrow \langle\pi, s', n', h', E[e]\rangle}
\hva \and
\inferrule[attr]{
  o' = {\tt getClassAttrObject}(o, x, h)
}
{\langle\pi, s, n, h, E[o.x]\rangle \rightarrow \langle\pi, s, c, h, E[o']\rangle}
\hva \and
\inferrule[new]{
    o_3 = {\tt getObject}(s, n, x) \\
    o_2 = {\tt getClassAttrObject}(o_3, {\tt \_\_init\_\_}, h) \\
}
{\langle\pi, s, n, h, E[\text{\primsem{new} $x$($y=o_1\dots$)}]\rangle \rightarrow \langle\pi, s, n, h, E[o_2(y=o_1\dots); o_3]\rangle}
\hva \and
\inferrule[attr-assign]{
    o_3 = {\tt getClassAttrObject}(o_1, x, h) \\
    \pi' = \pi[o_3 \rightarrow \pi(o_3) \cup \{o_2\}]
}
{\langle\pi, s, n, h, E[o_1.x := o_2]\rangle \rightarrow \langle\pi', s, n, h, E[o_3]\rangle}
\hva \and
\inferrule[import]{
    o_2 = {\tt getObject(s, m, x)} \\
    s' = {\tt addScope}(s, n, y, \primsem{var}) \\
    o_1 = (n, (y, \primsem{var})) \\
    \pi' = \pi[o_1 \rightarrow \pi(o_1) \cup \{o_2\}]
}
{\langle\pi, s, n, h, E[\text{\primsem{import} $x$ \primsem{from} $m$ \primsem{as} $y$}]\rangle \rightarrow \langle\pi', s', n, h, E[o_1]\rangle}
\hva \and
\inferrule[iter-iterable]{
    o' = {\tt getClassAttrObject}(o, {\tt \_\_next\_\_}, h) \\
}
{\langle\pi, s, n, h, E[\text{\primsem{iter} $o$}]\rangle \rightarrow \langle\pi, s, n, h, E[o'()]\rangle}
\hva \and
\inferrule[iter-generator]{
    {\tt getClassAttrObject}(o, {\tt \_\_next\_\_}, h) = \text{undefined} \\
}
{\langle\pi, s, n, h, E[\text{\primsem{iter} $o$}]\rangle \rightarrow \langle\pi, s, n, h, E[o()]\rangle}
\end{mathpar}}
\caption{Rules of the analysis.}\label{fig:semantics}
\vspace{-5mm}
\end{figure}

The analysis examines every expression
found in the intermediate representation
of the initial program,
and transitions the analysis state
according to the semantics of each expression.
The algorithm repeats this procedure until the state converges,
and the assignment graph is given
by the final state of the analysis.

Figure~\ref{fig:semantics} demonstrates
the state transition rules of our analysis.
The rules follow the form:
\[
{\langle\pi, s, n, h, E[e]\rangle \rightarrow \langle\pi', s', n', h', E[e']\rangle}
\]
In the following,
we describe each rule in detail.

According to the {\sc [e-ctx]} rule,
when we have an expression $e$ in the evaluation context $E$,
an assignment graph $\pi$,
a scope $s$,
a namespace $n$,
a class hierarchy $h$,
we can get an expression $e'$ in the evaluation context $E$,
if the initial expression $e$ evaluates to $e'$.
For what follows,
the binary operation $x \cdot y$ stands for
appending the element $y$ to the list $x$.

The {\sc [compound]} rule states
that when we have a compound expression
consisting of two objects $o_1$, $o_2$,
we return the last object $o_2$
as the result of the evaluation.
Observe that the evaluation of the compound expression
requires each sub-term to have been evaluated
to an object according to the evaluation contexts
shown in Figure~\ref{fig:syntax}.
The rest of the rules also follow
this behavior.

The {\sc [ident]} rule
describes the scenario when
the initial expression is an identifier $x$.
In this case,
the analysis retrieves the object $o$
corresponding to the identifier $x$,
in the namespace $n$,
based on the scope tree $s$.
To do so,
the analysis uses the function {\tt getObject($s, n, x$)},
which iterates every element $y$ of the namespace $n$
in the reverse order.
Then,
by examining the scope tree $s$,
it checks whether the element node $y$
has any child matching the identifier $x$.
In case of a mismatch,
the function {\tt getObject} proceeds
to the next element of the namespace.
Notice that the {\sc [ident]} rule
does not have any side-effect on the analysis state.

The {\sc [assign]} rule assigns
the object $o$ to the identifier $x$.
First,
the analysis adds the identifier $x$
in the current namespace $n$ of the scope tree $s$,
using the function {\tt addScope($s, n, x, \tau$)}.
This function adds an edge from the node
accessed by the path $n$
to the target node given by the definition $(x, \tau)$.
Second,
this rule updates the assignment graph
by adding an edge from the object
corresponding to the left-hand side of the assignment
(i.e., $o'$)
to that of the right-hand side (i.e., $o$).
This update says
that the variable $x$
defined in the namespace $n$ can point to the object $o$.

{\sc [func]} updates the scope tree.
In particular,
it adds the function $x$ to the current namespace $n$,
leading to a new scope tree $s'$.
Then,
it creates a new namespace $n'$
by adding the function definition $(x, \primsem{func})$
to the top of the current namespace.
It adds all function parameters,
and a virtual variable named {\tt ret}---which
stands for the variable holding the return value
of the function---to the newly-created namespace $n'$.
This results in a new scope tree $s^{(3)}$.
Finally,
the analysis proceeds to the evaluation of the body
of the function $x$ in the fresh namespace $n'$,
i.e., observe that the rule evaluates to $E[e]$.
The new namespace $n'$ correctly captures
that any variable defined in $e$,
is actually defined in the body of the function.

{\sc [return]} assigns the object $o$
to the virtual variable {\tt ret},
which is used for
storing the return value of a function
(recall the {\sc [func]} rule).
To do so,
the analysis updates the assignment graph
by adding a new edge from the object $o'$
corresponding to the return variable {\tt ret}
to the object $o$
which is the operand of~\prim{return}.
Finally,
this rule evaluates to the object $o'$
related to the return virtual variable {\tt ret}.

The inter-procedural flow is captured
by the {\sc [call]} rule.
Specifically,
when we encounter a call expression
$o_1(y=o_2\dots)$,
we examine the callee object $o_1$
associated with a function $f$
defined in a namespace $n'$.
Then,
the rule connects every parameter of $f$
with the appropriate argument passed
during function invocation (e.g.,
the counterpart object of the parameter
$y$ at call-site is $o_2$),
leading to a new assignment graph $\pi'$.
As an example,
consider again the graph of Figure~\ref{fig:ass_graph}.
The outgoing edges of the \{{\tt crypto.Crypto.apply, func}\} node
are created by this rule.
These edges imply that the parameter {\tt func} of
the {\tt crypto.Crypto.apply} function
may hold the functions {\tt cryptops.encrypt}
and {\tt cryptops.decrypt} passed
when calling {\tt crypto.Crypto.apply}
(Figure~\ref{fig:crypto-mod}).

The {\sc [class]} rule handles
class definitions.
The rule first adds the class $x$
to the scope tree through
the function {\tt addScope()},
and then gets every object
related to the base classes of $x$
(i.e., $y\dots$).
To achieve this,
the rule consults the scope tree
in the namespace $n$,
and gets a sequence of objects $t$
that respects the order in
which base classes are passed
during class definition.
We later explain why keeping
the registration order of base classes is important.
The rule then updates the class hierarchy
so that the freshly-defined class $x$
is a child of the base classes
pointed to by the identifiers $(y\dots)$.
After this,
the analysis works on the body of the class $e$
in a new namespace $n'$.
The new namespace contains the class definition
to the top of the current namespace
(i.e., $n \cdot (x, \primsem{cls})$).
Then,
the analysis starts examining
the body of the class
using the new namespace.

The {\sc [attr]} rule is similar to
{\sc [ident]}.
However,
this time,
in order to correctly retrieve the object
corresponding to the attribute $x$
of the receiver object $o$,
the analysis examines the hierarchy of classes $h$
through the function {\tt getClassAttrObject($o, x, h$)}.
This is the point where
our analysis is able to distinguish
attributes according to the location (i.e., $o$)
where they are defined.

To deal with multiple inheritance,
the function {\tt getClassAttrObject()}
respects the method resolution order
implemented in Python.
For example,
consider the following code snippet.
\begin{minted}[fontsize=\footnotesize,linenos,xleftmargin=20pt]{python}
class A:
  def func():
    pass

class B:
  def func():
     pass

class C(B, A):
   pass

c = C()
c.func()
\end{minted}
In the example above,
the method resolution order
is $C \rightarrow B \rightarrow A$,
because the class {\tt B}
is the first parent class of {\tt C},
while {\tt A} is the second one.
As a result,
{\tt c.func()} leads to the invocation of
function {\tt func} defined in class {\tt B},
as it is the first matching function
whose name is {\tt func}
in the method resolution order.
Correctly resolving class members
explains why
the domain of the class hierarchy
maps every object to a~\emph{sequence}
of objects rather than a set---we need to track the order
in which the parents
of a class are registered.

For object initialization,
we introduce the {\sc [new]} rule.
This rule gets the object $o_3$
associated with the definition of the class $x$.
Using the {\tt getClassAttrObject()} function,
the rule inspects the method resolution order of
the object $o_3$ to find the first
object $o_2$ matching the function {\tt \_\_init\_\_}.
Recall that this function is called
whenever a new object is created.
Observe how the~\prim{new} evaluates;
it reduces to $o_2(y=o_1\dots);o_3$.
That is,
we first call the constructor of the class
with the same arguments passed as in
the initial expression
(i.e., $o_2(y=o_1)$),
and then we return the object $o_3$
corresponding to the class definition,
which is eventually the result of the~\prim{new} expression.

The rule for attribute assignment $o_1.x := o_2$
describes the case
when the attribute $x$ is defined
somewhere in the class hierarchy of the
receiver object $o_1$.
In this case,
{\tt getClassAttrObject()}
returns the object $o_3$
associated with this attribute,
and the rule updates the assignment graph
so that $o_3$ points to the object $o_2$
from the right-hand side of the assignment.
If the attribute is not defined in the class hierachy,
(i.e., {\tt getClassAttrObject()} returns $\bot$)
the attribute assignment is similar to {\sc [assign]},
i.e.,
we first add the attribute $x$ to the current scope
through {\tt addScope()},
and then update the graph.
This case is omitted for brevity.

When we encounter an~\prim{import} $x$ \prim{from} $m$ \prim{as} $y$
expression,
we retrieve the object $o_2$ corresponding
to the imported identifier $x$,
which is defined in the module $m$.
Then,
we create an alias $y$ for $x$.
To do so,
we add $y$ to the scope tree of the current namespace,
and update the assignment graph
by adding an edge from
the object of $y$ to that of $x$.
Through this rule,
we are able to deal with Python's module system.

Consuming iterables and generators is supported
through the~\prim{iter} $x$ expression.
When the identifier $x$ points to an iterable,
(i.e., the object pointed to by $x$
has an attribute named {\tt \_\_next\_\_}),
we get the object $o'$ related to {\tt \_\_next\_\_}.
Then,
\prim{iter} evaluates to a call of $o'()$
(see the {\sc [iter-iterable]} rule).
If this is not the case,
we treat $x$ as a generator
({\sc [iter-generator]}).
In this case,
\prim{iter} reduces to a call of $x()$.
Recall from Section~\ref{sec:syntax}
that we model generators as thunks,
therefore this scenario
describes
the evaluation of these thunks
(generators)
when they are actually used (iterated).

{\bf Remark about analysis termination.}
The analysis traverses expressions,
and transitions the analysis state
based on the rules of Figure~\ref{fig:semantics},
until the state converges.
The analysis is guaranteed to terminate,
because the domains are finite.
Even in the presence of the domain of class hierarchy
$h \in ClassHier$
(Figure~\ref{fig:domains}),
which is theoretically infinite,
the analysis eventually terminates,
because a Python program cannot have
an unbounded number of classes.

\subsection{Call Graph Construction}
\label{sec:cg-construct}

\setlength{\textfloatsep}{4pt}
\begin{algorithm}[t]
\small
\caption{Call Graph Construction}
\label{alg:cg}
\DontPrintSemicolon
\SetKwInOut{Input}{Input}
\SetKwInOut{Output}{Output}
\Input{$p \in {\it Program}$\\
 $\sigma \in {\it State}$}
\Output{${\it cg} \in {\it CallGraph}$}
\SetKw{In}{in}
\SetKwFunction{getReachableFuns}{getReachableFuns}
\SetKwFunction{getObject}{getObject}
\ForEach{$e$ \In $\it Program$}{
    \While{$e \not\in {\it Obj}$}{
        $\langle\sigma, E[e]\rangle \rightarrow \langle\sigma', E[e']\rangle$\;
        \If(\tcp*[h]{Call Expression}){$e' = o_1(y=o_2\dots)$}{
            $(\pi, s, n \cdot f, h) \gets \sigma'$\;
            $c \gets \getReachableFuns{$\pi$, $o_1$}$\;
            $o_3 \gets \getObject{$s$, $n$, $f$}$\;
            ${\it cg} \gets {\it cg}[o_3 \rightarrow {\it cg}(o_3) \cup c]$ \tcp{Add Call Edges}
        }
        $e \gets e'$\;
    }
}
\Return{{\it cg}}
\end{algorithm}

After the termination of the analysis,
we build the call graph
by performing a final pass
on the intermediate representation
of the given Python program.
Algorithm~\ref{alg:cg}
describes the details of this pass.
The algorithm takes two elements
as input:
(1) a program $p \in {\it Program}$
of the model language 
whose syntax is shown in Figure~\ref{fig:syntax},
and (2) the final state $\sigma \in State$
stemming from the analysis step.
The algorithm produces a call graph:
\[
    {\it cg} \in {\it CallGraph} = {\it Obj} \hookrightarrow \mathcal{P}({\it Obj})
\]
The graph contains only objects
associated with functions.
An element $o \in {\it Obj}$ mapped to
a set of objects $t \in \mathcal{P}({\it Obj})$
means that the function $o$~\emph{may} call any function
included in $t$.

The algorithm inspects every expression $e$
found in the program (line 1),
and it evaluates $e$
based on the state transition rules
described in Figure~\ref{fig:semantics}.
The algorithm repeats the state transition rules,
until $e$ eventually reduces to an object (lines 2, 3).
Every time
when $e$ reduces to a call expression
of the form $o_1(y=o_2\dots)$ (line 4),
the algorithm gets the namespace where
this invocation happens
and retrieves the top element
of that namespace
(see $n \cdot f$, line 5).
After that,
the algorithm gets all functions
that the callee object $o_1$ may point to.
To do so,
it consults the assignment graph
through the function {\tt getReachableFuns($\pi, o_1$)},
which implements a simple Depth-First Search ({\sc dfs}) algorithm
and gets the set of functions $c$
that are reachable from the source node $o_1$.
In turn,
the algorithm updates the call graph ${\it cg}$
by adding all edges from
the top element
of the current namespace
to the set of the callee functions $c$
(lines 7, 8).
In other words,
the object $o_3$ (line 7)
representing the top element of the namespace,
where the call occurs,
is actually the caller of the functions
pointed to by the object $o_1$.

\subsection{Discussion \& Limitations}
\label{sec:approach:limitations}

One of our major design decisions is
to ignore conditionals and loops.
For instance,
when we come across an {\tt if} statement,
our analysis over-approximates the program's
behavior and considers both branches.
This design choice enables efficiency
without highly compromising
the analysis precision
(as we will discuss in Section~\ref{sec:evaluation}).
Other static analyzers~\cite{LWJC12,KDKW14,JMT09}
choose to follow a more heavyweight approach
and reason about conditionals.
These static analyzers,
though,
do not solely focus on
call-graph construction,
but rather they attempt to compute
the set of all reachable states
based on an initial one.
However,
for call-graph generation,
providing such an initial state
that exercises all feasible paths
(which is required in order to compute
a complete call graph),
especially when analyzing libraries,
is not straightforward.

In Python
where object-oriented features,
duck typing~\cite{duck},
and modules are extensively used,
it is important to separate attribute accesses
based on the namespace
where each attribute is defined.
This design choice boosts---contrary to prior work~\cite{FSSD13}---the
precision of our analysis
without sacrificing its scalability.

Our analysis does not fully support all of
Python's features.
First,
we ignore code generation schemes,
such as calls to the {\tt eval}
built-ins.
In general,
such dynamic constructs hinder
the effectiveness of any static analysis,
and dynamic approaches are often employed
as a countermeasure~\cite{gatekeeper,synode}.
Second,
our approach does not store
information about variables' built-in types, and
does not reason about the effects
of built-in functions.
Therefore,
attribute calls that depend on a specific built-in type
(e.g., {\tt list.append()}) are not resolved,
while the effects of functions
such as {\tt getattr} and {\tt setattr} are ignored.
Third,
we can only analyze modules
for which their source code
has been provided.
When a function---for which its code definition is not available---is called,
our method will add an edge to the function,
but no edges stemming from that function
will ever be added, and
its return value will be ignored.

\section{Implementation}
\label{sec:implementation}

We have developed~\pycg,
a prototype of our approach in Python~3.
For each input module,
our tool creates its scope tree and
its intermediate representation
by employing the
\textit{symtable}~\cite{symtable}
and \textit{ast}~\cite{ast}
modules respectively.

Our prototype discovers the file
locations of the different imported
modules to further analyze them by
using Python's \textit{importlib} module.
This is the module that Python uses internally
to resolve import statements.
We perform two steps.
First,
the file location of the imported
module is identified,
and then a \textit{loader} is used to import
the module's code.
In Python one can define custom
loaders for import statements,
which allowed us to use a loader
that logs the file locations discovered
and then exit without loading the code.
Then, in the second step,
our tool takes over and uses the
discovered file's contents to
iterate its intermediate representation in a recursive manner.
This allows us to resolve imports
in an efficient way.
Currently,
we only analyze discovered modules
that are contained in the package's namespace.

\section{Evaluation}
\label{sec:evaluation}

We evaluate our approach based on three research questions:
\begin{enumerate}[label={\bf RQ\arabic*}, leftmargin=2.1\parindent]
\item Is the proposed approach effective
in constructing call graphs for Python programs?
(Sections~\ref{sec:evaluation:micro}
 and~\ref{sec:evaluation:macro})
\item How does the proposed approach
stand in comparison with existing
open-source,
static-based
approaches for Python?
(Sections~\ref{sec:evaluation:micro}
	and~\ref{sec:evaluation:macro})
\item What is the performance of our approach?
    (Section~\ref{sec:evaluation:performance})
\end{enumerate}

\noindent
Further,
we show a potential application through
the enhancement of GitHub's ``security advisory''
notification service.

\subsection{Experimental Setup}
\label{sec:evaluation:setup}

We use two distinct benchmarks:
(1) a micro-benchmark suite containing
112 minimal Python programs, and
(2) a macro-benchmark suite of five
popular real-world Python packages.
We ran our experiments on a Debian 9 host
with 16 {\sc cpu}s and 16 {\sc gb}s of {\sc ram}.

\subsubsection{Micro-benchmark Suite}
\label{sec:evaluation:micro-benchmark}

We propose a test suite
for benchmarking call graph
generation in Python.
Based on this suite,
researchers can evaluate and compare
their approaches against a common standard.
Reif et al.~\cite{RKEH19}
have provided a similar suite for Java,
containing unique call graph test cases,
grouped into different categories.

Our suite consists of 112 unique and
minimal micro-benchmarks that cover a wide
range of the language's features.
We organize our micro-benchmarks
into 16 distinct categories,
ranging from simple function calls to
more complex features such as
twisted inheritance schemes.
Each category contains a number of tests.
Every test includes (1) the source code,
(2) the corresponding
call graph (in {\sc json} format),
and (3) a short description.
Categorizing and adding a
new test is relatively easy.
The source code of each test
implements only a single execution path
(i.e., no conditionals and loops)
so there is a straightforward correspondence to its call graph.
Table~\ref{tbl:test-suite-items}
lists the categories along with
the number of benchmarks they incorporate
and a corresponding description.

\begin{table}[tb]
\centering
\caption{Micro-benchmark suite categories.}
\label{tbl:test-suite-items}
\resizebox{\linewidth}{!}{%
\begin{tabular}{lrl}
\hline
{\bf Category} & {\bf \#tests} & {\bf Description} \bigstrut\\
 \hline
{\tt parameters}        & 6     & Positional arguments that are functions \\
{\tt assignments}       & 4     & Assignment of functions to variables \\
{\tt built-ins}         & 3     & Calls to built in functions and data types \\
{\tt classes}           & 22    & Class construction, attributes, methods \\
{\tt decorators}        & 7     & Function decorators \\
{\tt dicts}             & 12    & Hashmap with values that are functions \\
{\tt direct calls}      & 4     & Direct call of a returned function ({\tt func()()}) \\
{\tt exceptions}        & 3     & Exceptions \\
{\tt functions}         & 4     & Vanilla function calls \\
{\tt generators}        & 6     & Generators \\
{\tt imports}           & 14    & Imported modules, functions classes \\
{\tt kwargs}            & 3     & Keyword arguments that are functions \\
{\tt lambdas}           & 5     & Lambdas \\
{\tt lists}             & 8     & Lists with values that are functions \\
{\tt mro}               & 7     & Method Resolution Order ({\sc mro}) \\
{\tt returns}           & 4     & Returns that are functions \\
\hline
\end{tabular}}
\end{table}

{\bf Addressing Validity Threats}:
The internal validity of the
micro-benchmark suite depends on
the range of Python features
that it covers.
To address this threat,
we presented the suite to two researchers,
who have professionally worked as Python developers
(other researchers have applied similar
methods to verify their work~\cite{RPW19}).
Then,
we asked them to rank the suite
(from~1 to~10) based on the following criteria:
\begin{enumerate}
\item Completeness: Does it cover all Python features?
\item Code Quality: Are the tests unique and minimal?
\item Description Quality: Does the description adequately
describe the given test case?
\end{enumerate}

\noindent
The first reviewer provided
a~9.7 ranking in all cases.
The second indicated an excellent
(10) code and description quality but
ranked lower (6) the completeness of the
benchmarks.

Both reviewers provided corresponding feedback.
In their comments,
they suggested some code cleanups
and asked for more comprehensive descriptions on
some complex benchmarks.
Regarding the completeness of the suite,
they pointed out missing tests
for some common features such as built-in functions
and generators.
We applied the reviewers' suggestions by
refactoring the affected benchmarks and
improving their descriptions.
Furthermore,
we implemented more tests for
some of the missing functionality.

\subsubsection{Macro-benchmarks}
\label{sec:evaluation:macro-benchmark}

We have manually generated call graphs
for five popular real-world packages.
The packages were chosen as follows.
First,
we queried the {G}it{H}ub {\sc api} for
Python repositories sorted by their number of stars.
Then,
we downloaded each repository and counted the
number of lines of Python code.
If the repository contained less
than 3.5k lines of Python code,
we kept it.
Table~\ref{tbl:project-descriptions} presents
the GitHub repositories we chose along with
their lines of code,
GitHub stars and forks,
together with a short description.

Currently,
there is no acceptable implementation
generating Python call graphs in an
effective manner, so
the first author manually inspected the
projects and generated their
call graphs in {\sc json} format,
spending on average~10 hours for each project.
We opted for medium sized projects
(less than 3.5k LoC), so
that we could minimize
human errors.
To further verify the validity of
the generated call graphs,
we examined the output
of \pycg\, {\it Pyan}, and {\it Depends} and
identified~90 missing edges from a total of 2506.

\begin{table}[tb]
\centering
\caption{Macro-benchmark suite project details.}
\label{tbl:project-descriptions}
\resizebox{\linewidth}{!}{%
\begin{tabular}{lrrrl}
\hline
{\bf Project} & {\bf LoC} & {\bf Stars} & {\bf Forks} & {\bf Description}\bigstrut \\
\hline
{\tt fabric}                   & \nnum{3236} & 12.1k & 1.8k & Remote execution \& deployment \\
{\tt autojump}                  & \nnum{2662} & 10.8k & 530  & Directory navigation tool \\
{\tt asciinema}             & \nnum{1409} & 7.9k  & 687  & Terminal session recorder \\
{\tt face\_classification}   & \nnum{1455} & 4.7k  & 1.4k & Face detection \& classification \\
{\tt Sublist3r}              & \nnum{1269} & 4.4k  & 1.1k & Subdomains enumeration tool \\
\hline
\end{tabular}}
\end{table}

\subsection{Micro-benchmark suite results}
\label{sec:evaluation:micro}

\begin{table}[tb]
\centering
\caption{Micro-benchmark results for \pycg\ and {\it Pyan}. {\it Depends} is unsound in all cases and complete in 110/112 cases and is omitted.}
\label{tbl:benchmark-comparison}
\begin{tabular}{lrrrr}
\hline
{\bf Category} & \multicolumn{2}{c}{\bf PyCG} & \multicolumn{2}{c}{\bf Pyan}\bigstrut \\
& {\bf Complete} & {\bf Sound} & {\bf Complete} & {\bf Sound} \bigstrut \\
\hline
{\tt assignments}       & 4/4   & 3/4   & 4/4   & 4/4 \\
{\tt built-ins}         & 3/3   & 1/3   & 2/3   & 0/3 \\
{\tt classes}           & 22/22 & 22/22 & 6/22  & 10/22 \\
{\tt decorators}        & 6/7   & 5/7   & 4/7   & 3/7 \\
{\tt dicts}             & 12/12 & 11/12 & 6/12  & 6/12 \\
{\tt direct calls}      & 4/4   & 4/4   & 0/4   & 0/4 \\
{\tt exceptions}        & 3/3   & 3/3   & 0/3   & 0/3 \\
{\tt functions}         & 4/4   & 4/4   & 4/4   & 3/4 \\
{\tt generators}        & 6/6   & 6/6   & 0/6   & 0/6 \\
{\tt imports}           & 14/14 & 14/14 & 10/14  & 4/14 \\
{\tt kwargs}            & 3/3   & 3/3   & 0/3   & 0/3 \\
{\tt lambdas}           & 5/5   & 5/5   & 4/5   & 0/5 \\
{\tt lists}             & 8/8   & 7/8   & 3/8   & 4/8 \\
{\tt mro}               & 7/7   & 5/7   & 0/7   & 2/7 \\
{\tt parameters}        & 6/6   & 6/6   & 0/6   & 0/6 \\
{\tt returns}           & 4/4   & 4/4   & 0/4   & 0/4 \\
\hline
{\bf Total} & {\bf 111/112} & {\bf 103/112} & {\bf 43/112} & {\bf 36/112} \bigstrut\\
\hline
\end{tabular}
\end{table}

The benchmarks included in
the micro-test suite have a limited scope
and are designed to cover
specific functionalities
(such as decorators and lambdas).
Table~\ref{tbl:benchmark-comparison}
lists the results of our evaluation.
For each benchmark belonging to
a specific category,
we show if our prototype and {\it Pyan}
generated complete or sound call graphs.
Note that a call graph is complete when
it does not contain any call edges that
do not actually exist (no false positives),
and sound when it contains every
call edge that is realized
(no false negatives).

\pycg\ produces
a complete call graph in almost all
cases (111/112).
In addition,
it produces sound call graphs for
103 out of 112 benchmarks.
The lack of soundness
is attributed to not fully covered functionalities,
i.e., Python's starred assignments.

{\it Pyan} produces either complete
or sound call graphs at a much lower rate.
However,
for assignments,
{\it Pyan} turns out as a more sound method
because it supports them in a better manner.
We performed a qualitative analysis
on the call graphs generated by {\it Pyan} to check
the reasons behind its performance.
We observed that {\it Pyan} produces incomplete
call graphs because it creates call edges
to class names as well as their {\tt \_\_init\_\_} methods
(see also Section~\ref{sec:background:limitations}).
Also it generates imprecise results
because it does not support all
of Python's functionality,
($0/6$ generators and $0/3$ exceptions),
ignores the inter-procedural flow of functions
($0/6$ parameters and $0/4$ returns),
misses calls to imported ones ($4/14$),
and fails to support classes ($10/22$).

The evaluation of {\it Depends}
shows both its fundamental
strengths and limitations.
Recall that each benchmark implements
a single execution path
and includes a call coming from the module's namespace.
Our results indicate that {\it Depends} does not
identify calls from module namespaces,
and therefore soundness is never achieved (0/112).
In terms of completeness,
{\it Depends} achieves an almost perfect score (110/112)
due to its conservative nature---i.e., it adds an edge when it has
high confidence that it will be realized.

\subsection{Macro-benchmark results}
\label{sec:evaluation:macro}

By using our macro-benchmark,
we have examined the three tools
in terms of
precision and recall.
Precision measures
the percentage of
valid generated calls over
the total number of generated calls.
Recall measures
the percentage of valid generated calls
over the total number of calls.
To do so,
we manually generated the call graphs
of the examined packages.

Table~\ref{tbl:oracle-comparison}
presents our results.
The missing entries for {\it Pyan}
indicate that the tool crashed
during the execution.
Our findings show that \pycg\
generates high precision call graphs.
On all cases,
more than $98\%$ of the
generated call edges are true positives,
while on one case
none of the generated call edges
are false positives.
Recall results show
that on average,
$69.9\%$ of all call edges are successfully retrieved.
The missing call edges
are attributed to the approach's
limitations
(recall Section~\ref{sec:approach:limitations}),
and missing support
for some functionalities.

{\it Pyan} shows average precision and low recall.
{\it Pyan}'s average precision appears
because the tool adds call edges
to class names instead of just their {\tt \_\_init\_\_} methods.
Also,
it does not track the inter-procedural
flow of functions,
which is the reason why it has low recall.
For instance,
the implementation of the
{\tt face\_classification} package
mostly depends on functions declared
in external packages.
{\it Pyan} ignores such calls which in turn
leads to a 7.6\% recall.

Finally,
{\it Depends} shows high precision (98.7\%)
and low recall.
The high precision of {\it Depends}
can be attributed to its conservative nature.
Furthermore,
{\it Depends} does not track higher order functions and
does not include calls coming
from module namespaces.
This in turn,
leads to its low recall.

\begin{table}[tb]
\centering
\caption{Macro-benchmark results and tool comparison.}
\label{tbl:oracle-comparison}
\vspace{-1mm}
\resizebox{\linewidth}{!}{%
\begin{tabular}{lrrrrrr}
\hline
{\bf Project} & \multicolumn{3}{c}{\bf Precision (\%)}
& \multicolumn{3}{c}{\bf Recall (\%)} \bigstrut\\
& {\bf PyCG} & {\bf Pyan} & {\bf Depends} & {\bf PyCG} & {\bf Pyan} &
 {\bf Depends} \bigstrut\\
\hline
{\tt autojump}                      & $99.5$  & $66.5$      & $99.2$  &  $68.2$     & $28.5$       & $22.5$ \\
{\tt fabric}                        & $98.3$  & -           & $100$   &  $61.9$     & -            & $6.3$ \\
{\tt asciinema}                     & $100$   & -           & $98.1$  &  $68$       & -            & $15.5$ \\
{\tt face\_classification}          & $99.5$  & $86.8$      & $96.2$  &  $89.7$     & $7.6$        & $5.7$ \\
{\tt Sublist3r}                     & $98.8$  & $69.8$      & $100$   &  $61.6$     & $25.6$       & $21.9$ \\
\hline
{\bf Average} & {\bf 99.2} & {\bf 74.4} & {\bf 98.7} & {\bf 69.9} & {\bf 20.6} & {\bf 14.4} \bigstrut\\
\hline
\end{tabular}}
\end{table}

\subsection{Time and Memory Performance}
\label{sec:evaluation:performance}

\begin{table}[tb]
\centering
\caption{Time and memory comparison.}
\label{tbl:time-memory-comparison}
\vspace{-0.5mm}
\resizebox{\linewidth}{!}{%
\begin{tabular}{lrrrrrr}
\hline
{\bf Project}
& \multicolumn{3}{c}{\bf Time (sec) } & \multicolumn{3}{c}{\bf Memory (MB)} \bigstrut \\
& {\bf PyCG} & {\bf Pyan} & {\bf Depends} & {\bf PyCG} & {\bf Pyan} & {\bf Depends} \bigstrut \\
\hline
{\tt autojump}                      & 0.76  & 0.42    & 2.37  & 62.7  & 37.8 & 27.1 \\
{\tt fabric}                        & 0.77  & -      & 1.83  & 60.9  & -    & 18.5 \\
{\tt asciinema}                     & 0.87  & -      & 2     & 61.6  & -    & 19.4 \\
{\tt face\_classification}          & 0.92  & 0.38   & 2.49  & 60.9  & 35.3 & 25.6 \\
{\tt Sublist3r}                     & 0.51  & 0.33   & 2.01  & 60    & 35.8 & 19.4 \\
\hline
{\bf Average} & {\bf 0.77} & {\bf 0.38} & {\bf 2.14} & {\bf 61.2} & {\bf 36.3} & {\bf 22} \bigstrut \\
\hline
\end{tabular}}
\end{table}

We use the macro-benchmark suite
as a base for our time and memory evaluation.
Table~\ref{tbl:time-memory-comparison}
presents the time and memory performance
metrics of the three tools.
The execution time was calculated
using the {\sc unix} \textit{time} command,
while the memory consumption was measured
using the {\sc unix} \textit{pmap} command.
The metrics presented
are the average
out of 20 runs.

The results show that {\it Pyan}
is more time efficient, and
that {\it Depends} is more memory efficient.
\pycg\ and {\it Pyan} generate a call graph
for the programs in the benchmark
($\leq$~3.5k LoC)
in under a second,
while {\it Depends} requires more
than two seconds on average.
Furthermore,
all tools use a
reasonable amount of memory,
with \pycg, {\it Pyan} and {\it Depends} using
on average $\sim$61.2,
$\sim$36.3 and $\sim$22{\sc mb}s
of memory respectively.
Overall,
\pycg\
is on average~2 times slower than {\it Pyan},
and uses~2.8 times the
amount of memory that {\it Depends} uses.
We attribute the differences
in execution time between {\it Pyan} and \pycg\
to the fact that
{\it Pyan} performs two passes
of the {\sc ast} in comparison
to \pycg\ performing a
fixpoint iteration (Section~\ref{sec:approach}).
{\it Depends} is overall slower,
because it spends most of its execution time
parsing the source files.
In terms of memory,
{\it Pyan} and {\it Depends} store less information
about the state of the analysis
leading to better memory performance.

\subsection{Case Study: A Fine-grained Tracking of Vulnerable Dependencies}
\label{sec:github}

GitHub sends a notification to
the contributors of a repository when
it identifies a dependency to
a vulnerable library.
However,
this notification
does not indicate if the project
invokes the function containing
the defect.
We show that \pycg\
can be employed to enhance the
service with method-level information
that may further
warn the contributors.

To highlight the usefulness of our
method in this context,
we performed the following steps.
First we accessed GitHub's
``Advisory Database''~\cite{gadv}.
Then,
we searched for vulnerable Python packages
sorted by the severity of the defect.
In many occasions the accompanying {\sc cve}
(Common Vulnerabilities and Exposures) entries
did not include further details about the defects.
We disregarded such instances and focused on
the first two cases that provided information
about the functions that contained the vulnerability:
(1) {\sc p}y{\sc yaml}~\cite{pyyaml}
(versions before 5.1),
a {\sc yaml} parser affected by
{\sc cve}-2017-18342~\cite{cve18342},
and
(2) Paramiko~\cite{paramiko}
(multiple versions before 2.4.1),
an implementation of the {\sc ssh}v2 protocol affected by
{\sc cve}-2018-7750~\cite{cve7750}.
Both packages were imported by
thousands of projects,
9226 for {\sc p}y{\sc yaml}
and 1097 for Paramiko.
We could not clone all dependent repositories
because some were private and others
did not exist any more:
we managed to download 570 {\sc p}y{\sc yaml}
and 322 Paramiko dependent projects.
Then,
we ran our tool on each project and generated
corresponding call graphs for
106 out of the 570 {\sc p}y{\sc yaml} dependent projects
and 76  out of the 322 Paramiko dependent projects---%
the projects that \pycg\ failed to generate call graphs were written in Python~2.
Finally,
we queried the generated call graphs
to check if the vulnerable functions were included.
We found that the 
vulnerable function in {\sc p}y{\sc yaml} (i.e., {\tt load})
was invoked by 42/106 projects.
In Paramiko we found
that the problem method ({\tt start\_server})
was not utilized at all by any of the 76 projects.
We also observed that 12 projects did not invoke any
library coming from Paramiko.
Paramiko was needlessly included in the
requirement files of the dependents.
That was not a false negative from our part:
we manually checked that \pycg\ did not miss any invocation.

\section{Related Work}
\label{sec:related}

{\bf Call Graph Generation.}
Methods that generate call graphs can be
either dynamic~\cite{XN02},
or static~\cite{MNGL98}.
Dynamic approaches
usually produce fewer false positives,
but suffer from performance issues.
Also,
they are able to analyze a single
execution path,
and their effectiveness relies on
the program's input.
Static approaches are
more time efficient and
can typically cover a wider range of execution paths,
trying to capture all possible program's behaviors.
Several approaches~\cite{EKS01,GFFS17,LLTX17},
try to combine the two
so they can get improved results.

There are plenty of methods
and tools targeting call graph generation
for statically-typed programming languages
such as Java.
{\sc doop}~\cite{BS09} and
~{\sc wala}~\cite{WALA12}
follow a context-sensitive,
points-to analysis method.
{\sc paddle}~\cite{LH08},
a similar approach,
employs Binary Decision Diagrams ({\sc bdd}s)~\cite{bdd}.
Finally, {\sc opal}~\cite{EKHR18}
is a lattice-based approach
written in Scala.
Ali et al.~\cite{AO12},
implement {\sc cgc},
a partial call graph generator for Java,
with the main focus being efficiency.
They ignore calls coming from
externally imported libraries, and
only analyze the source code of
a given package.
We are currently following a similar approach,
but we aim to efficiently analyze
external dependencies in the future.

Moving to dynamic languages,
Ali et al.~\cite{ALLL19}
convert Python source code into
{\sc jvm} bytecode, and
use the existing implementations
for Java~\cite{WALA12,VCGH10,LH03}
to generate its call graph.
However,
they argue that generating precise
call graphs using this method is infeasible,
and sometimes the output has more than
$96\%$ of false positives.
{\it pycallgraph}~\cite{PCG19}
generates Python call graphs
by dynamically analyzing
one execution path.
Thus the analysis is not practical
and one should pair it
with another method (e.g., fuzzing)
to retrieve meaningful results.
On the {J}ava{S}cript front,
Feldthaus et al.~\cite{FSSD13}
implement a flow-based approach for
the generation of call graphs.
They evaluate against
call graphs generated by
a dynamic approach paired with instrumentation,
achieving
$\geq 66\%$ precision and
$\geq 85\%$ recall.
Other {J}ava{S}cript
call graph generators include,
{\sc ibm wala}~\cite{WALA12},
{\sc npm} call graph~\cite{G19},
Google closure compiler~\cite{BOC10},
Approximate Call Graph ({\sc acg})~\cite{FSSD13}, and
Type Analyzer for JavaScript ({\sc tajs})~\cite{JMT09}.
{\sc tajs} implements
a lattice-based flow-sensitive approach
using abstract interpretation.
Although, such an approach yields
more promising results,
it comes with a performance cost.

{\bf Call Graph Benchmarking and Comparison.}
Reif et al. present Judge~\cite{RKEH19},
a toolchain for analyzing
call graph generators for Java.
At its core,
the toolchain contains
a test suite with
benchmarks for a range of
Java features.
The authors then
proceed to compare
Java call graph generators,
namely
Soot~\cite{VCGH10,LH03},
{\sc wala}~\cite{WALA12},
{\sc doop}~\cite{BS09} and
{\sc opal}~\cite{EKHR18}.
Sui et al.~\cite{SDER18},
also present a
test suite of Java benchmarks, and
they use it to evaluate and compare
Soot~\cite{VCGH10,LH03},
{\sc wala}~\cite{WALA12}, and
{\sc doop}~\cite{BS09}.
The above benchmark suites are very similar,
leading to Judge consolidating them
into one benchmark suite.
Recall
our very similar implementation of
a micro-benchmark suite
from Section~\ref{sec:evaluation:setup}.

{\bf Static Analysis for Dynamic Languages.}
Numerous advanced frameworks aim
for the static analysis of
JavaScript programs.
{\sc safe}~\cite{LWJC12}
provides a
formally specified
static analysis framework
with the goal of being
flexible, scalable and pluggable.
{\sc jsai}~\cite{KDKW14}
is a formally specified 
and provably sound platform
using abstract interpretation.

Other JavaScript approaches
target different aspects
of its functionality.
Madsen et al.
implement {\sc radar}~\cite{MTL15}
a tool that identifies bugs in
event-driven JavaScript programs.
Sotiropoulos et al.~\cite{async}
propose an analysis targeting
asynchronous functions.
Bae et al.~\cite{BCLR14},
implement {\sc safe}\textsubscript{{\sc wapi}}
a tool aimed
at identifying possible {\sc api} misuses.
Park et al.~\cite{PWJR15} propose
{\sc safe}\textsubscript{{\sc wa}pp},
a static analyzer for
client-side JavaScript.

Fromherz et al.~\cite{FOM18}
implement a prototype that
soundly identifies run-time errors
by evaluating the
data types of Python variables
through abstract interpretation.
In comparison,
our approach does not infer
the data types of variables and
focuses on the generation of call graphs.

\section{Conclusion}
\label{sec:conclusion}

We have introduced
a practical static approach
for generating Python call graphs.
Our method performs
a context-insensitive
inter-procedural analysis
that identifies the
flow of values
through the construction of a graph
that stores
all assignment relationships
among program identifiers.
We used two benchmarks to
evaluate our method, namely
a micro- and a macro-benchmark suite.
Our prototype showed high rates of
both precision and recall.
Also,
our micro-benchmark suite can serve
as a standard for the evaluation of
future methods.
Finally,
we applied our approach in a
real-world case scenario,
to highlight how it can aid
dependency impact analysis.\\

\noindent
{\bf Acknowledgments.}
We thank the anonymous reviewers
for their insightful comments and
constructive feedback.
This work has received funding
from the European Union's Horizon 2020
research and innovation programme
under grant agreement No. 825328.

\bibliographystyle{IEEEtran}
\bibliography{pycg}

\end{document}